\documentclass[useAMS,usenatbib,usegraphicx]{mn2e}

\def\hmpc {\,h^{-1}\,{\rm Mpc}} 
\def\hunit {\, {\rm km \, s^{-1} \,Mpc^{-1} } } 
 
\def\zeq {z_{\rm eq}} 
 
\def\neff{ N_{\rm eff}} 
\def\xrad{ X_{\rm rad}} 

\def\simlt{\mathrel{\lower0.6ex\hbox{$\buildrel {\textstyle <} 
  \over {\scriptstyle \sim}$}}}

\def\simgt{\mathrel{\lower0.6ex\hbox{$\buildrel {\textstyle >}
 \over {\scriptstyle \sim}$}}}

\def\newtwo { }

\title[The BAO amplitude and $N_{eff}$ ]
{On the baryon acoustic oscillation amplitude as a probe of 
  radiation density }

\author[Will Sutherland \& Lukasz Mularczyk] 
{Will Sutherland$^{1}$\thanks{E-mail: w.j.sutherland@qmul.ac.uk} 
 \& Lukasz Mularczyk$^{1}$
\\
$^{1}$School of Physics and Astronomy, Queen Mary University of London, 
  Mile End Road, London E1 4NS, UK 
}  

\begin{document}

\date{MNRAS: Accepted 2013 December 13. Received 2013 December 12; 
  in original form 2013 July 4}

\pagerange{\pageref{firstpage}--\pageref{lastpage}} \pubyear{2014}

\maketitle

\label{firstpage}

\begin{abstract}

 The baryon acoustic oscillation (BAO) feature in the distribution of galaxies 
 has been widely studied as an excellent standard ruler for probing cosmic
 distances and expansion history, and hence dark energy. In contrast, 
  the amplitude of the BAO feature has
 received relatively little study, mainly due to limited 
 signal-to-noise, and complications due to galaxy 
  biasing, effects of non-linear clustering and dependence on 
  several cosmological parameters.  
 As expected, the amplitude of the BAO feature
  is sensitive to the cosmic baryon fraction: 
 {\newtwo for standard radiation content, 
  the cosmic microwave background (CMB) 
 acoustic peaks constrain this precisely and the BAO amplitude is 
   largely a redundant cross-check.  However, 
  the CMB mainly constrains the redshift
 of matter-radiation equality, $\zeq$, and the baryon/photon ratio: if a
  non-standard radiation density $(\neff)$ is allowed, 
   increasing $\neff$ while matching the CMB peaks 
  leads to a reduced baryon fraction and a lower relative BAO amplitude. 
 We construct an observable for the relative area of the BAO feature
 from the galaxy correlation function  (Eq.~\ref{eq:wb}); 
  from linear-theory models, we find that this is  
  mainly sensitive to $\neff$ and quite insensitive to other
   cosmological parameters. 
  More detailed work from N-body simulations will be needed to 
   constrain the effects of non-linearity and scale-dependent galaxy bias 
  on this observable. }   

\end{abstract}


\section{Introduction}

The detection of baryon acoustic oscillations (BAOs) in the 
 large-scale distribution of galaxies in both the SDSS \citep{eis05} and 
 2dFGRS \citep{cole05} redshift surveys was a major milestone for 
 cosmology, strongly supporting the standard paradigm for
 structure formation based on gravitational instability including 
  cold (or warm) dark matter. 
 Recently, there have been several new independent 
 measurements of the BAO feature in galaxy redshift surveys, e.g. from 
 SDSS-DR8 \citep{perc10}, WiggleZ \citep{blake11}, 6dFGRS \citep{beutler11}, 
 an angular measurement from SDSS-DR9 \citep{seo12}, and 
  a first measurement from BOSS \citep{anderson12}, 
   which are all consistent with
 the concordance $\Lambda$CDM model at the few-percent level. 

 The BAO feature in galaxy clustering \citep{peeb-yu70, be84, eh98, mwp99} 
  has a very similar origin  to the acoustic peaks in the 
 cosmic microwave background (CMB) temperature power spectrum.  
 Most recent attention has focused on the length-scale of the BAO 
 feature, used as a standard ruler to measure cosmic distances
 in units of the sound horizon $r_s(z_d)$ at the baryon drag
 epoch. 
  Many theoretical and computational 
  studies \citep{seo08,seo10} have concluded that
 the comoving length-scale of the BAO feature evolves by
  $\sim 0.5\,\%$ between the CMB era and the recent past $z \sim 0.3$ 
  due to the non-linear growth of structure, but this 
  shift can be corrected down to the $0.1\,\%$ level using 
   reconstruction methods \citep{esss07,pad12}. 
 Therefore,  the BAO feature is probably the best-understood standard
  ruler in the moderate-redshift universe, and  
 when combined with CMB observations it offers 
  great power for probing the cosmic expansion history
  and therefore the properties of dark energy \citep{wein12}. 
 {\newtwo These 
 BAO distance measurements are complementary to those 
 from type-Ia supernovae, have potentially smaller 
  systematic errors, and can offer
 direct information on the time-variable $H(z)$ without differentiation.
  On the downside, cosmic variance sets a floor on the BAO 
 precision in the low-redshift universe, $\sim 1\,$percent at
 $z \sim 0.25$ and worsening below this \citep{se07} 
}. 

 However, in this paper we look at a different property, specifically 
 the overall amplitude of the BAO feature, rather than the
 length-scale.  As expected, the amplitude is mainly
 sensitive to the cosmic baryon fraction (relative to total matter), 
  $f_b$. 
 Until now, the BAO amplitude has received much less attention than
 the length-scale, for two main reasons: 
  firstly, recent CMB results from 
 WMAP \citep{wmap9}, SPT \citep{keis11, story12} and ACT \citep{act} 
  measure the baryon fraction to around 4 percent  
 relative precision (given standard assumptions), while
 the strongest detection of the BAO peak \citep{anderson12} 
  gives $\sim 6\sigma$ significance
  or $\sim 16\%$ error in amplitude.  Secondly, complications
 due to galaxy bias, the non-linear growth of structure, 
  redshift-space distortions
  and the uncertain global shape of the power spectrum 
  make it challenging to extract the baryon fraction from the BAO feature,
 even with very large future redshift surveys.  

 However, we note that parameter estimates from the CMB 
 are subject to a significant degeneracy between $f_b$ and
  the total radiation density in the CMB era, 
 usually parametrized by an effective number of neutrino 
  species $\neff$.  
 {\newtwo Recent reviews of $\neff$ are given by e.g. \citet{riemer13} and 
 \citet{abaz12}.
} 
   
 In contrast, the BAO 
 feature is sensitive to the baryon fraction
 rather directly: therefore, combining CMB measurements (primarily sensitive
 to the physical baryon density
 $\omega_b$ and the redshift of matter-radiation equality 
  $\zeq$) with
 a BAO measurement {\newtwo sensitive to} 
 $f_b$ may provide an interesting probe of 
 the radiation density, $\neff$. This may be less precise than
  other methods, but is largely {\newtwo complementary}. 

The plan of the paper is as follows: in \S~\ref{sec:bao}
 we review the main effects of varying cosmological
 parameters, including $f_b$ and radiation density, 
   on CMB and BAO observations.  
In \S~\ref{sec:measfb} we present numerical predictions of the BAO
 feature for a set of models (selected to give a good match to
 WMAP) with varying matter density and 
 $\neff$, and we derive a statistic based on the galaxy correlation
  function which is sensitive to $f_b$ and $\neff$, but cancels
  galaxy bias and dark energy to leading order.   
 We summarize our conclusions in \S~\ref{sec:conc}.  
 Most of this work was completed before the {\em Planck} 
 release in March 2013,
 so we mainly use WMAP-9 fit parameters \citep{wmap9} 
  as our baseline. The adjustments
 post-{\em Planck} are moderate, and we discuss
 the implications of recent 
 {\em Planck} results (Planck Collaboration, \citet{planck-xvi}) 
  in \S~\ref{sec:planck}. 

 Throughout the paper we use the standard notation that
 $\Omega_i$ is the present-day density of species $i$ relative to the
 critical density; and the physical density $\omega_i \equiv \Omega_i h^2$, 
 with $h \equiv H_0 / (100 \hunit)$. 


\section{BAOs, radiation density $\neff$ and the cosmic baryon fraction} 
\label{sec:bao} 

\subsection{Overview of BAOs} 

 The BAO feature appears as a single hump in the galaxy correlation
 function $\xi(r)$, or equivalently a series of decaying wiggles
  in the power spectrum (see \cite{esw07} for a 
  clear explanation in real space, and \cite{bh10} and \cite{wein12} 
 for reviews).  
 The length-scale of the hump is very close to 
  the comoving sound horizon $r_s(z_d)$
   at the baryon drag epoch $z_d \simeq 1020$; this $z_d$ 
  is commonly defined by the fitting formula 
  Eq.~4 of \citet{eh98}. {\newtwo (This formula is defined for
 standard $\neff$; however the dependence of $z_d$ on $\neff$ is 
   weak, so the error from adopting the fitting formula is small). }  
  This comoving length is predicted precisely
  mainly from CMB constraints, and several very large redshift
 surveys are ongoing or planned to exploit this as a standard ruler 
  to measure cosmic distances at $0.2 \simlt z \simlt 2.5$ and 
  thus probe dark energy.   

 Standard cosmological models contain a density of
 collisionless dark matter $\sim 5\times$ larger than 
 the baryon density.  This explains naturally 
  why the acoustic peaks in the CMB
  power spectrum have large relative amplitude, while the 
  BAO feature is relatively weak in the late-time  
   galaxy correlation function.  
  Qualitatively, this occurs because the acoustic peaks 
  at last scattering appeared only in the power spectrum of baryons, not dark 
   matter: 
   the peaks are prominent in the CMB 
  because almost all CMB photons last scattered 
  off a free electron. After decoupling, the distribution of baryons and 
  dark matter became averaged together by gravitational growth 
  of structure over the next few $e$--foldings
  between the CMB era and redshift $z \sim 20$ \citep{esw07},
   well before the formation 
   of large galaxies. 
  The dark matter dominates in this averaging, so
  the BAO signal in the galaxy correlation function 
   becomes diluted by a factor $\sim f_b$.   
 In the following we define the baryon fraction as 
 \begin{equation} 
 f_b \equiv {\omega_b \over \omega_c + \omega_b} 
 \end{equation} 
 so that the denominator includes CDM and baryons, but excludes 
  massive neutrinos.

 Thus, the BAO peak amplitude provides a potential measure of the baryon
 fraction; this has been used as a simple and compelling argument 
 against MOND-type modified gravity theories 
  without non-baryonic dark matter \citep{dod11}.  
 
 We can make this more quantitative and use the BAO feature
  to directly estimate the cosmic baryon fraction. However, there
  are several reasons why this has received little attention to date:
 \begin{enumerate} 
 \item  
 Recent observations of the CMB power spectrum \citep{wmap9} 
  measure the physical baryon density $\omega_b$ to 
 $\approx 2$ percent precision, and (assuming standard $\neff$), 
  measure the physical matter density $\omega_m$ to 3 percent;  
  the errors on these are weakly correlated, so 
  this gives an estimate of the cosmic baryon fraction $f_b$ to 
  $\le 4$ percent relative precision.  
 (Recent results from the {\em Planck} mission have
   improved this to the $\sim 2$ percent level, see \S~\ref{sec:planck}).   
\item The BAO feature is affected by several effects: 
  it is blurred by the non-linear growth of structure, and amplified
  by galaxy bias, which is challenging to measure and 
   may also be scale-dependent. 
\item The overall large-scale shape of the galaxy power spectrum
   also depends on other cosmological parameters, and this will also have
   some influence on the BAO peak shape.  
\end{enumerate} 

Thus, at first sight it appears that BAOs cannot compete in precision 
 with the CMB as a probe of $f_b$.  This is partly true, but with 
 an important caveat: the CMB-based estimates of $f_b$ are significantly
 degenerate with the total radiation density in the CMB era.   

\subsection{Radiation density} 
\label{sec:rad} 

At this point we define our notation on densities: 
 as usual, we define the parameter $\neff$ 
 such that the radiation density at matter-radiation equality is 
\begin{equation} 
\label{eq-neff} 
  \rho_{rad} = \rho_\gamma (1 + 0.2271 \, \neff)
\end{equation} 
 where $\rho_{rad}$, $\rho_\gamma$ are densities of (total) 
 radiation and photons
  respectively. Here $\neff$ is an ``effective'' number of neutrinos, 
  but in fact it is not specific to neutrinos and 
  counts any species (except photons) which were relativistic until
 around matter-radiation equality. 
 Assuming the standard population of only three very light neutrinos with 
  the oscillation parameters given by solar, atmospheric
 and beam-based neutrino experiments, the
  value of $\neff$ can be accurately predicted as $\neff = 3.046$ 
 \citep{mangano05}; here the additional $0.046$ arises from 
  a small residual coupling of neutrinos to baryons and photons 
  at the epoch of electron/positron annihilation. 

 It is also convenient to define the scaled radiation 
  density $\xrad$ by  
\begin{equation}
\label{eq-xrad} 
  \xrad \equiv {\rho_{rad} \over 1.6918 \, \rho_\gamma } 
  =  1 + 0.134 \,(\neff - 3.046) 
 \end{equation} 
 {\newtwo where the factor of $1.6918$ is the bracket in 
 Eq.~\ref{eq-neff} for}
 $\neff = 3.046$; therefore $\xrad = 1.00$ for standard 
  radiation content, and for example
   $\xrad = 1.134$ for $\neff = 4.046$, i.e. the case of 
  additional ``dark radiation'' with 
  energy density equal to one standard neutrino flavour.  
 Here $\neff$ and $\xrad$ are equivalent, but the latter is convenient 
 later since several parameters of interest scale almost 
  as half-integer powers of $\xrad$. 
 
 There are now several known routes to probe $\neff$ from observations: 
  historically it was first constrained by big bang nucleosynthesis, 
  \citep{ssg77, mangano11}. 
  However, $^4$He is the nuclide with the main sensitivity to $\neff$, 
  and observational measurements of the primordial $^4$He
   abundance, $Y_P$, appear to be limited by systematic errors;
  over the past 25 years the estimates of $Y_P$ have shifted significantly 
 upwards,  but the realistic error bars have not much improved.
  Unless a new better method of measuring $Y_P$ can be found, we
   cannot expect dramatic progress from the Helium route. 
 Recently, a constraint on $\neff$ has been derived from 
  deuterium abundance \citep{pc12}, but this currently 
 relies on only a single object, and also uses the baryon density
  derived from the CMB.  

  Secondly, the CMB damping tail at high multipoles 
   $\ell \simgt 1400$ is sensitive to $\neff$ 
  \citep{jkks96, bs03, hou11};  
  and several recent measurements (\citealt{hou12};\citealt{riemer13};
  \citealt{planck-xvi}) 
  give tantalizing but not decisive hints
  for a value higher than the standard $3.046$. 
  However, the CMB damping tail method is significantly 
  degenerate with other possible new parameters, including running of
  the spectral index $n_s$ and non-standard helium abundance
  $Y_P$ \citep{hou11, joudaki13}, so other {\newtwo complementary}
  probes of $\neff$ are desirable. 

 Thirdly, combining CMB data with a direct local 
  measurement of $H_0$ can also probe
  $\neff$; however, using CMB+$H_0$ alone 
  is critically dependent on other 
   assumptions such as $w = -1$ and flatness. An improvement on 
  the $H_0$ method is given by \citet{suth12}, 
  who showed that a {\em theory-free} measurement of $r_s$ can be 
 obtained by  combining a low-$z$ BAO redshift survey 
  and a suitable absolute distance measurement 
  to a matched redshift (specifically $4 \overline{z} /3$, 
  where $\overline{z}$ is the characteristic redshift of the BAO 
  survey).  This almost cancels the distance effects  
   from dark energy and curvature; comparing such a direct $r_s$ 
  measurement to CMB data (which mainly constrains $r_s \sqrt{\xrad}$
  rather than $r_s$ alone)  therefore probes $\neff$.  
  The above-mentioned method 
 is less theory-dependent than the CMB damping tail, 
  but requires a challenging measurement of a distance to 
   $z \sim 0.3$ to $\sim 2$ percent absolute accuracy. 
 
 We will demonstrate 
  below that the amplitude of the BAO feature provides a fourth
 possible probe of $\neff$:  this is currently much less precise than the
 known methods above,   but involves different assumptions 
  and systematics; with future massive redshift surveys expected 
 in the next decade,  it may provide a useful complement to the 
 better-known methods above.  

\subsection{Cosmological parameter set} 
\label{sec:pars} 

 The present-day photon density $\omega_\gamma = \Omega_\gamma h^2$ 
 is very well constrained by the observed CMB temperature \citep{fixsen} 
 and spectrum to be $\omega_\gamma \simeq 
 1/40440$; and we define  $\omega_{cb} \equiv \omega_c + \omega_b$ 
  to be the physical matter density today, 
  specifically CDM and/or WDM plus baryons, excluding 
  neutrinos. 
 Defining $\zeq$ as the redshift of matter-radiation equality, and
  simply assuming that the photon density scales with redshift 
  as $\propto (1+z)^4$, 
  and matter conservation so CDM and baryon densities
  scale $\propto (1+z)^3$ 
 (i.e. no decaying dark matter, dark energy to dark matter
   transitions, or other exotic effects) leads to the following  identities:  
 \begin{equation} 
 \label{eq:omcb}  
  \omega_{cb} = { (1+\zeq) \, \xrad \over 23904 } 
 \end{equation} 
 \begin{equation} 
 \label{eq:hzeq} 
    h =  \sqrt{ { (1+\zeq) \, \xrad \over 23904 \; \Omega_{cb} }  } 
\end{equation} 
\begin{equation} 
 \label{eq:fb} 
  f_b \equiv {\omega_b \over \omega_{cb} } = 
   { 23904 \; \omega_b \over (1+\zeq) \, \xrad } 
\end{equation} 
 The equations above are independent of assumptions about dark energy and
 flatness. They remain valid for the case of small non-zero neutrino mass,  
  $m_\nu \simlt 0.3 \, {\rm eV}$: since our definition of $\Omega_{cb}$ 
   excludes the contribution from 
  massive neutrinos today, while neutrinos this light were
  almost fully relativistic at the era of matter-radiation equality. 
  This assumption is reasonable given recent upper limits on 
 neutrino mass from CMB+galaxy clustering data (\citealt{planck-xvi}; 
 \citealt{giusarma13}; \citealt{riemer13b}). 
 Clearly, low-mass neutrinos are matter-like at
 $z \simlt 100$ and do contribute to $\Omega_m$ in late-time
  observables, but we treat $\Omega_{\nu}$ as a separate 
 contribution. 

 We now choose a basic 6+1 set of cosmological 
 parameters as 
\begin{equation} 
\label{eq:pars} 
 \zeq ; \ \Omega_{cb}; \ \omega_b; \ A; \ n_s; \ \tau; \ \xrad
\end{equation} 
   where the first three and $\xrad$ are
 defined above, as usual 
  $A$ is the scalar perturbation amplitude (which cancels in the following),
 $n_s$ is the scalar spectral index 
  and $\tau$ is the optical depth to last scattering. 
 Then, $h$, $\omega_{cb}$ and $f_b$ are 
  derived parameters from Eqs.~(\ref{eq:omcb})--(\ref{eq:fb}). 
 We may add optional parameters, 
 curvature $\Omega_k$, dark energy equation of state $w$ 
 and present-day neutrino density $\Omega_\nu$ 
  defaulting to $0, -1, \approx 0.0013$ respectively (for minimal 
   neutrino mass).  
 Then $\Omega_{tot} \equiv 1 - \Omega_k$, and the 
  dark energy density $\Omega_{DE}$ is
  another derived parameter, via 
  $\Omega_{DE} = 1  - \Omega_{cb} - \Omega_k - \Omega_\nu$.  

  This parameter set (\ref{eq:pars})
   including $\zeq$ and $\Omega_{cb}$ in the basic six 
   looks unconventional compared to the more common
  choice including $\omega_m, \, \Omega_{DE}$ as two of the basic parameters; 
  but for variable $\neff$, our set links more naturally to 
   observational constraints
   as we will see below; see also Appendix A, 
  and the discussion in Section~4.2 of \citet{suth12}.  
 To summarize the latter, dimensionless observables such as the CMB 
 acoustic wavenumber $\ell_*$,\footnote{Here, following WMAP
 convention, $\ell_* \equiv \pi / \theta_*$, 
 with acoustic angle $\theta_* \equiv r_S(z_*) / (1+z_*) D_A(z_*)$ and
  $z_*$ is the redshift of decoupling.}
  and distance ratios from BAO and SNe provide good constraints on dimensionless
 parameters including $\zeq$ and $\Omega_{cb}$; but there remains
  one overall scale degeneracy between $\neff$ and 
   dimensionful parameters such as $H_0$, $t_0$, $\rho_{cb}$. 
  (Parameters such as $h$, $\omega_{cb}$ 
   are only pseudo-dimensionless since they are relative to
  an arbitrary choice of $100 \hunit$, and these are affected by this
  degeneracy).  

  It has been  shown by several authors (\citealt{hs96, jkks96, 
   bs03, jvpk04, komatsu11}) 
 that the heights of the first few acoustic peaks in the CMB primarily 
  constrain the redshift of
  matter-radiation equality $\zeq$, {\em not} the 
   physical matter density $\omega_{cb}$.\footnote{Strictly, it is
  the ratio $(1+\zeq)/(1+z_*)$ which is important in the CMB, 
  where $z_*$ is the decoupling redshift;  however in practice
  the relative uncertainty in $z_*$ is much 
  smaller than in $\zeq$, so we ignore this for simplicity.}  
  These latter two parameters are equivalent 
   if we force $\neff \simeq 3.04$, but if we allow $\neff$ to be free
   they are no longer equivalent,
    and then $\zeq$ is constrained much better than $\omega_{cb}$ by CMB data 
  (see e.g. \citealt{komatsu11}).    

 The WMAP data also constrains the baryon density $\omega_b$ accurately. 
  The effect of baryons on the CMB derives mainly from the 
  baryon/photon ratio; given the photon density measured very accurately
  by COBE \citep{fixsen},  
  the $\omega_b$ estimate from WMAP is only weakly dependent on  
   $\neff$ or $\xrad$. 

 Measuring both $\omega_b$ and $\zeq$ immediately 
  gives us the product $f_b \, \xrad$ from Eq.~\ref{eq:fb};  so, 
 the key point from the above is that the first few CMB acoustic peaks
  provide an accurate constraint on the product $f_b \, \xrad$, 
  but $f_b$ and $\xrad$ are significantly degenerate.  
 Therefore, adding a {\newtwo non-CMB} observable which is sensitive to $f_b$ 
  can provide another probe of $\neff$. 

  In the next section, we show that the relative amplitude of
 the BAO peak in galaxy clustering may provide such a test: it 
   depends mainly on $f_b$, with weak 
  sensitivity to other parameters. Thus, comparing a BAO-based estimate
  of $f_b$ to a CMB-based measurement (approximately $f_b \xrad$)  
 can provide a new probe of the radiation 
 density which is largely independent of existing methods.
   A strong point of this method is that the 
  CMB can measure $\zeq$ using only the first 
  three acoustic peaks; for models near concordance parameter values, 
  the ratio of the third to first
  peak height is especially sensitive to $\zeq$ (\citealt{hfzt01, page03}), 
 and the third peak at $\ell \approx 800$ is only weakly
 affected by Silk damping which dominates at $\ell \simgt 1500$.  
 Thus, while we need CMB data at $\ell < 1000$, 
 this method is not strongly dependent on the CMB damping tail 
 and other possible early-time nuisance parameters.


\section{ Estimating baryon fraction from the BAO peak } 
\label{sec:measfb} 

\subsection{The BAO equivalent width} 
\label{sec:wb} 

We noted above that the CMB power spectrum from WMAP
  constrains $f_b \xrad$ to $\sim 4$ percent, 
 (and this improves to $\sim 2\%$ with {\em Planck} data); therefore, 
 an estimate of $f_b$ derived from the BAO amplitude can translate into
 a probe of $\xrad$ or equivalently $\neff$.  

 However, deriving $f_b$ from the BAO feature is affected
 by several complications listed below:
 firstly there is galaxy bias, which may be scale-dependent.  
 Here we choose to work with the correlation function rather than the
 power spectrum, since the former  
 makes the BAO feature a single hump which simplifies the analysis.  
  In the  linear-bias approximation, 
  the galaxy and matter correlation functions are related by 
  $\xi_{gg}(r) \simeq b^2 \xi_{mm}(r)$, therefore
  a suitable ratio of correlation functions near the BAO bump
  vs outside the bump can cancel galaxy bias if it is
  scale-independent.  This is believed to be a good approximation
  at linear scales $k < 0.1\, h \, {\rm Mpc}^{-1}$ 
 \citep{angulo08,baugh13},
 but a better understanding of galaxy formation may be required to 
  clarify this. 

 Secondly, the BAO bump is blurred by the non-linear growth of structure, 
  mainly due to peculiar motions \citep{esss07};
 this both lowers its height and broadens its shape, and 
 causes a small shift in central position. However, 
  it is shown by \citet{ow11} 
 that non-linear growth almost conserves the total area of the bump; 
 thus, measuring the bump {\em area} rather than its height
  is relatively insensitive to the non-linear growth of structure.  

 Thirdly, the global shape of $\xi(r)$, with $r$ 
  in $\hmpc$ units, depends on other quantities 
  including $\Omega_m$ and dark energy 
 equation of state,  which are not tightly constrained by the CMB.  
 However, we show later that if we define $u = r / r_s$ to be
 the ratio of comoving separation $r$ 
  to the sound horizon scale, then the broad-band shape of
  $\xi(u)$ on intermediate scales depends 
  mostly on $\zeq$: 
  nearly all shape dependence on other parameters is collapsed 
  into $\zeq$, which is already well determined by the CMB. 

 Therefore, we define the following observable $W_b$ 
 from the measured galaxy correlation function $\xi_{gg}$, as
  a measure of the BAO ``equivalent width'': 
  we define 
\begin{equation} 
\label{eq:wb} 
  W_b \equiv  { \int_{u_3}^{u_4} u^2 
  \left[ \xi_{gg}(u) - \xi_{nb}(u) \right] \, du 
    \over \int_{u_1}^{u_2} u^2 \, \xi_{gg}(u) \, du } 
\end{equation} 
where $\xi_{gg}(u)$ is the observed galaxy correlation function 
 in units of $u \equiv r / r_b$, $r_b$ is the bump scale (here the
 value of $r$ at the peak in $r^2 \xi_{gg}(r)$), and  
  $\xi_{nb}(u)$ is a smooth ``no bump'' curve, 
  here a polynomial fitted to the regions of
 $\xi_{gg}(u)$ just outside the BAO bump. Then $u_1, \ldots, u_4$
 are arbitrary dimensionless limits of integration, where  
  $u_3, u_4$ span almost the full area of the bump; while 
 $u_1 , u_2$ are intermediate scales non-overlapping
  with the bump, used for normalization. 
 There is a compromise here, since 
 we want to avoid the non-linear regime $u_1 \simlt 0.15$, while
  at $u_1 \simgt 0.6$ the measurement noise in $u^2 \xi(u)$ generally 
  increases, and 
  becomes more sensitive to systematic errors.   
 In the following we choose $u_1 = 1/3, u_2 = 2/3$ as simple values which 
  give a well-measured signal 
  in the linear regime, and are not too far below the bump scale
  to minimise the possible effects of scale-dependent bias.  

 In the above definition, a constant bias in $\xi_{gg}$ cancels
 out as long as it is scale-independent on large scales 
  $r \simgt 30 \hmpc$;  
 while a multiplicative stretch of cosmic distance
 scales (e.g. from varying dark energy) also cancels in $W_b$, since we 
 are measuring at fixed fractions of the comoving scale $r_b$ which 
  is fitted from the data.  
 The ratio between $r_s$ and the horizon size at matter-radiation
 equality is determined almost entirely by $\zeq$, so 
  we expect $W_b$ defined as above to be mainly sensitive to the baryon 
  fraction $f_b$ as desired. 

 To verify this and test parameter dependences,   
  we next evaluate $W_b$ from the linear matter power spectrum 
   for some example theoretical models 
  generated by CAMB.\footnote{We used the 2011 January release of CAMB,
 by A. Lewis and A. Challinor,  available from {\tt http://camb.info/} } 
 
\subsection{Dependence of $W_b$ on $\neff$ and $\zeq$} 
\label{sec:camb} 

 Here, we evaluate $W_b$ (defined above) for a set of six representative 
  models which are all consistent with CMB data up to 2012.  
 All models are flat $\Lambda$CDM ($\Omega_{tot} = 1$, $w = -1$), 
 and have $n_s$ fixed to 0.96  
 and $\omega_b = 0.0226$ in accordance with WMAP. 
 We vary $\zeq$ and $\neff$, and also adjust $\Omega_{cb}$ to preserve
  the CMB acoustic scale $\ell_*$.  

\begin{table*} 
\begin{tabular}{|l|c|c|c|c|c|c|c|} 
\hline 
Model  & $\zeq$ & $\Omega_{cb}$ &  $\neff$ &  $\omega_{cb}$ & $f_b$ & $H_0$ 
   & $W_b$ \\ 
       &        &               &          &             &    &  ($\hunit$)  
  &  \\ 
\hline
L3  & 3101 &  0.247  &  3.04  &  0.1298 & 0.174 & 72.5  & 0.612 \\ 
L4  & 3101 &  0.247  &  4.04  &  0.1471 & 0.154 & 77.1  & 0.561 \\ 
{\bf C3}  & {\bf 3264} &  {\bf 0.279}  &  {\bf 3.04}  &  
  {\bf 0.1366} & {\bf 0.165} & {\bf 70.0}  & {\bf 0.614} \\ 
C4  & 3264 &  0.279  &  4.04  &  0.1549 & 0.146 & 74.5  & 0.561 \\ 
H3  & 3428 &  0.315  &  3.04  &  0.1434 & 0.158 & 67.5  & 0.608 \\ 
H4  & 3428 &  0.315  &  4.04  &  0.1626 & 0.139 & 71.9  & 0.560 \\ 
\hline
\end{tabular}
\caption{Cosmological parameters for the six example 
  models discussed in the text. 
 All models have $\Omega_{tot} = 1$, $w = -1$ and $\omega_b = 0.0226$.  
  Model C3 (bold) is our baseline, while model C4 has $\neff = 4.04$ but
   unchanged $\zeq$ and $\Omega_{cb}$.  
 Models labelled L and H have $\zeq$ forced respectively 5\% lower/higher
 than C, then $\Omega_{cb}$ adjusted to preserve the CMB 
 acoustic scale.  Values of $\omega_{cb}$, $f_b$ and $H_0$ are
  derived from the first three.  The last column gives the 
  value of $W_b$ as defined in Eq.~\ref{eq:wb}, calculated by
  integrating the linear-theory matter correlation function. } 
\label{tab:models}
\end{table*} 

  For our ``base'' model (hereafter C3) we set 
 $\neff = 3.046$, $\zeq = 3264$,  $\Omega_{cb} = 0.279$; 
  therefore  $\omega_{cb} = 0.1366$ and $h = 0.700$.  
 For model C4 we add a fourth light neutrino species, 
  but retain identical values of $\zeq$ and $\Omega_{cb}$; 
   therefore C4 has $\omega_{cb}$ and $h$ 
  increased by $13.4\%$ and $6.5\%$ respectively relative to C3. 
  (Here $\omega_b$ is held at 0.0226 for both models, so 
  model C4 has dark matter density $\omega_c$
    increased by slightly more than 13.4\%, while $f_b$
 is reduced by a factor of 0.882).    

The overall shape of the correlation function also depends 
 significantly on $\zeq$: to explore this dependence, we choose
 two models (labelled L3, L4) with $\zeq$ fixed to 5\% lower than C3, 
 and respectively $\neff = 3.04$ and $4.04$; likewise 
  another two models (H3, H4) with $\zeq$ fixed 5\% higher than C3.  
 For these models, $\Omega_{cb}$ is adjusted in order to preserve the 
 CMB acoustic scale $\ell_* \equiv \pi / \theta_*$.  The resulting 
  parameter values are shown in Table~\ref{tab:models}. 
 Since these models are all flat, 
 they do not quite follow the CMB geometrical 
  degeneracy, but they do follow the related 
  degeneracy of constant $\ell_*$ or horizon angle as outlined
  in \cite{psp02}. 

\begin{figure*}
\includegraphics[angle=-90, width=15cm]{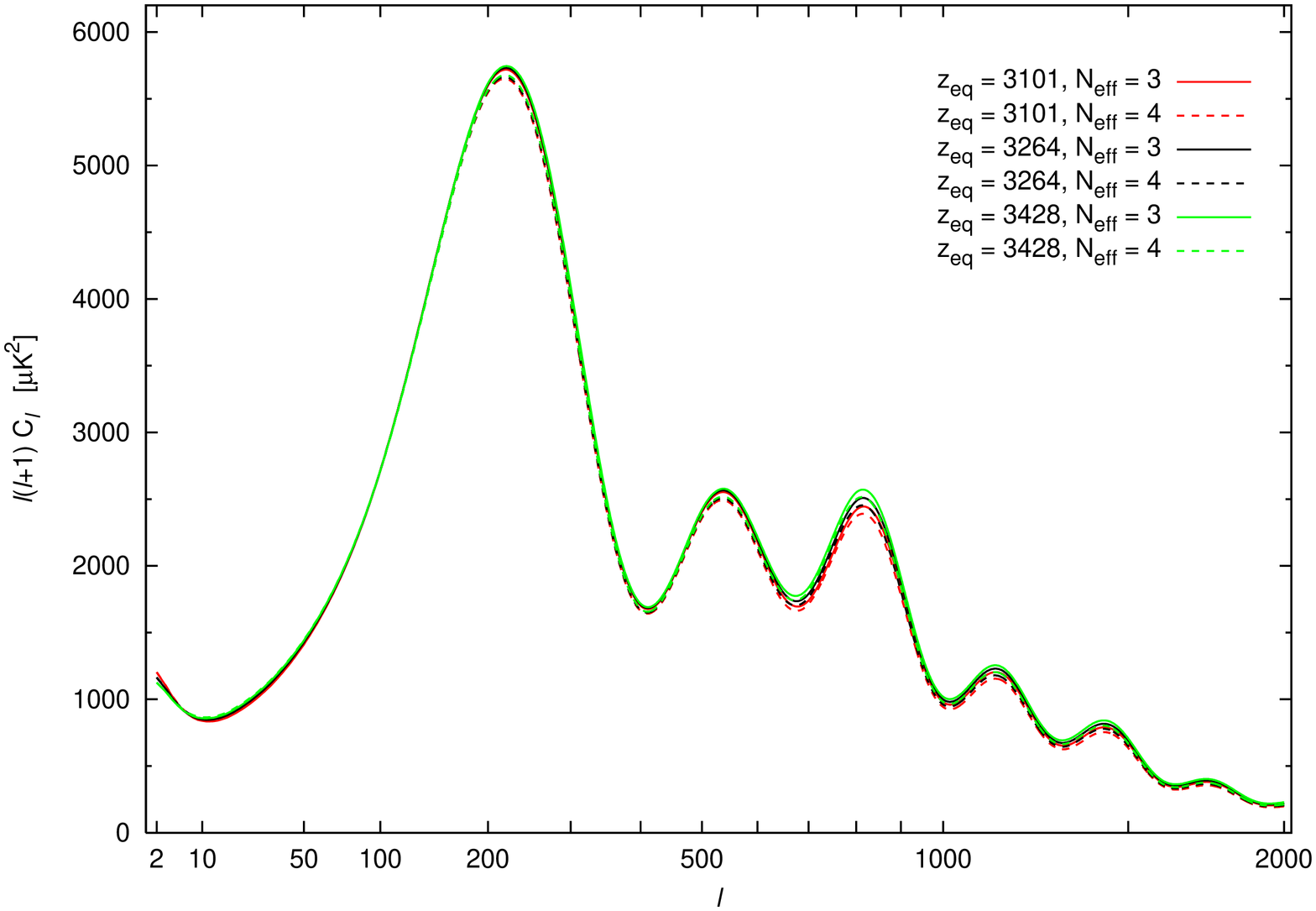} 
\caption{ 
  This figure shows the CMB temperature power spectra for the six 
 example models from Table~\ref{tab:models}; all are normalized 
  to match model C3 at $\ell = 100$. The horizontal 
  axis is linear in $\sqrt{\ell}$ for improved resolution 
  at low $\ell$.   Models with 
 $\neff = 3.04$ are solid lines; models with $\neff = 4.04$ are
 dashed lines.  The values of $\zeq$ are labelled. 
\label{fig:cmb} } 
\includegraphics[angle=-90, width=15cm]{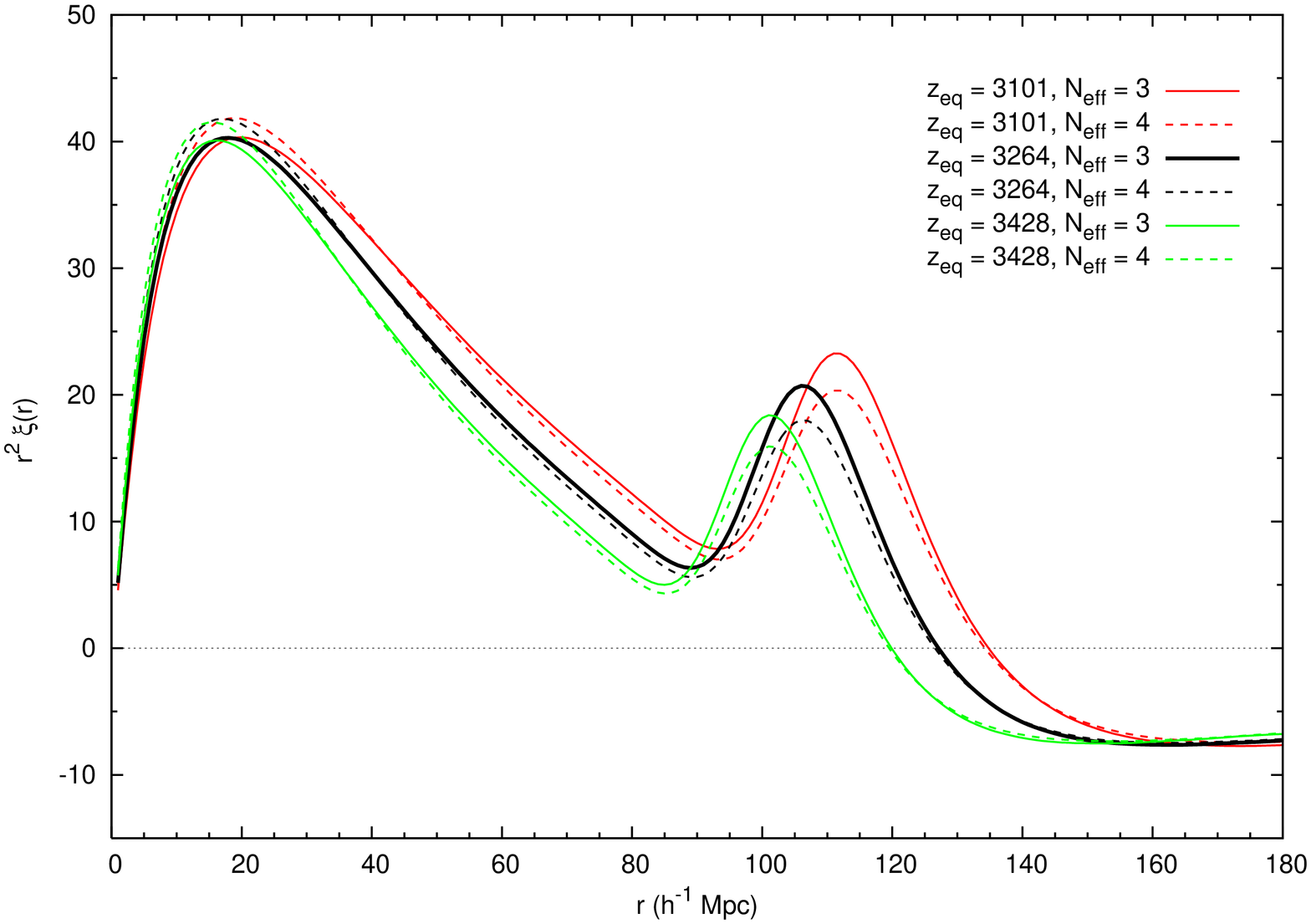} 
\caption{ 
  This figure shows the linear-theory matter
 correlation function for the six 
 example models from Table~\ref{tab:models}; 
  the ordinate is $r^2 \xi_{mm}(r)$ for clarity.  
 Models with 
  $\neff = 3.04$ are solid lines; models with $\neff = 4.04$ are
 dashed lines. The thick solid line is model C3 ; the L and H models
  are respectively higher/lower at $r \sim 50 \hmpc$.  
\label{fig:xi} } 
\end{figure*}

 We used CAMB to evaluate 
  the CMB temperature power spectra for the above six models;
 these are shown
  in Figure~\ref{fig:cmb}, normalized to match model
 C3 at $\ell = 100$. 
 Clearly the CMB spectra  
  are very similar for all our models, since the acoustic
 scales are matched by construction, and the variations in $\zeq$ are
 only $\pm 5$ percent. 
  Minor differences are apparent,  notably around the
  third peak (which is positively correlated with $\zeq$), while the
   effects of $\neff$ appear mainly in the damping tail and are
   small at $\ell < 1000$.  
 We repeat here that $\omega_b = 0.0226$ and $n_s = 0.96$ have been held fixed
 in all models for simplicity, in order to highlight the effects of
  $\zeq$ and $\neff$.  Clearly, allowing $n_s$ and $\omega_b$  
  to float to fit CMB data would
 result in model spectra that are even more similar, especially
  if a running spectral index is also allowed. 

 We took the linear-theory matter power spectra for the above six
 models generated by CAMB, and then Fourier transformed them
 to obtain the real-space matter correlation functions; these  
  are shown in Figure~\ref{fig:xi}, with the
  $x-$axis in units of $\hmpc$ corresponding
 to the observable from a low-$z$ redshift survey.   
 
 For the matter correlation functions in Figure~\ref{fig:xi}, 
  the differences between models are much more obvious than in the CMB: 
  the position of the BAO bump (in $\hmpc$ units) is 
  insensitive to $\neff$ for fixed $\zeq, \Omega_{cb}$, but it 
  does shift with $\zeq$.  
  In fact, as explained in Appendix~A,
 the BAO bump location is more sensitive to $\Omega_{cb}$
  than $\zeq$; but changing 
  $\zeq$ required us to adjust $\Omega_{cb}$ to conserve 
 the CMB acoustic scale, and it is actually the change
  in $\Omega_{cb}$ which dominates the shift of the bump location.  
 The other notable feature in Figure~\ref{fig:xi} is that
 all the $\neff = 4.04$ models have a slightly reduced BAO peak amplitude, 
  as qualitatively expected given their smaller $f_b$.  

\begin{figure*} 
\includegraphics[angle=-90, width=18cm]{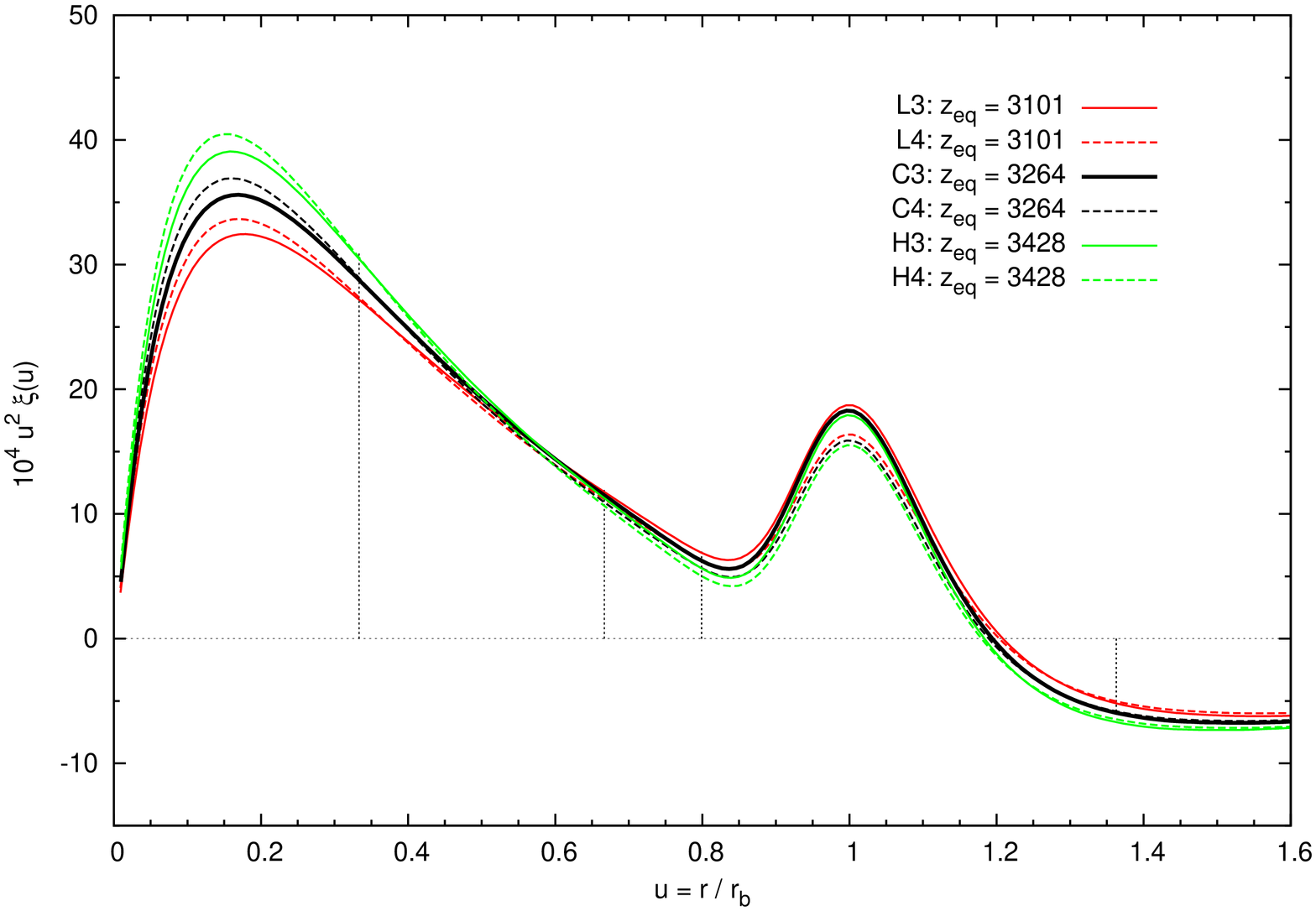} 
\caption{ 
  This figure shows the linear-theory matter
 correlation functions for the six 
 example models from Table~\ref{tab:models}, now with the $x-$axis 
  scaled so that $u = r / r_b$ and the BAO bump is at $u = 1$.  
  The ordinate is $10^4 u^2 \xi(u)$.  
  Models with 
  $\neff = 3.04$ are solid lines; models with $\neff = 4.04$ are
 dashed lines. The thick solid line is model C3 ; the H and L models
  are respectively higher/lower around $u \sim 0.2$.  Vertical dotted
 lines illustrate chosen integration limits $u_1$, $u_2$, $u_3$, $u_4$ 
  as used in Eq.~(\ref{eq:wb}). } 
\label{fig:xi_scaled}
\end{figure*} 

 To highlight the effects of varying $\neff$, in 
  Figure~\ref{fig:xi_scaled} 
  we plot the matter correlation functions as a function of
  $u = r/r_b$, so that the BAO bump
 appears at $u = 1$. 
   This Figure shows clearly
 that the bump amplitude is {\newtwo mainly sensitive to } $\neff$, while the 
  broad-band shape (the ratio of power at $u \sim 0.2$ to that at 
  $u \simgt 0.6$) 
  is governed mainly by $\zeq$. This is understandable since the 
 broad-band shape is determined  by the scale of the {\newtwo turnover}  
  in the matter  power spectrum, which is {\newtwo directly
   proportional} to the particle horizon size at $\zeq$. 
  This scale in observable $\hmpc$ 
  units depends on several cosmological parameters. 
  However, as noted in e.g. Eq.~B2 of \cite{suth12}, the 
   {\em ratio} of the BAO sound horizon $r_s(z_d)$ 
  to the particle horizon $r_H$ at $\zeq$ (both in comoving units) 
  has a simpler dependence: this ratio is well approximated by simply
 \begin{equation} 
 \label{eq:rhrat} 
  {r_s(z_d) \over r_H(\zeq)} \simeq 
   1.275 \left( 1+\zeq \over 3201 \right)^{0.75} \ ;
\end{equation} 
  since the sound speed $c_s(z)$
  is well constrained by the WMAP baryon density. The dependence
 on other parameters such as $\Omega_{cb}$, $h$, $\xrad$ is almost 
  entirely compressed into $\zeq$, and 
 the ratio is completely independent of late-time parameters such 
 as $w$, $\Omega_k$.  
  Thus the changes in $\xi(u)$ in Figure~\ref{fig:xi_scaled}
 are largely driven by the differing  $\zeq$ and $f_b$ 
 between the six models, and adding optional parameters such as
  $\Omega_k$, $w$ will have minimal effect. 
 
Here it is also noteworthy that the zero-crossing in $\xi(u)$ occurs close
 to $u \sim 1.2$ for all the models; this offers an interesting
  possible consistency
 test for the $\Lambda$CDM framework which is largely
  insensitive to galaxy bias. However, this 
  is observationally challenging to measure since the 
 zero-crossing is much more sensitive than the BAO bump 
 position and amplitude to broad-band systematic errors in 
 the observed $\xi(u)$.  
 
 The denominator of Eq.~\ref{eq:wb} is mainly sensitive to 
  the broad-band large-scale power at $k \simlt 0.1 \, h \, {\rm Mpc}^{-1}$, 
  which as above depends on the {\newtwo turnover} scale in the 
  matter power spectrum. 
  If we measured this in a fixed
  range of Mpc or $\hmpc$, this would depend on quantities such as
  $\Omega_{cb}$ and $w$, which would seriously 
    degrade our ability to measure $f_b$; but since
  we chose our mid-scale power estimate as a fixed {\em fraction} of
  the BAO length rather than a fixed range in $\hmpc$, 
  this mostly cancels the dependence on low-redshift
  parameters such as $\Omega_{cb}$, $\Omega_k$ and $w$; the
  broad-band shape of $\xi(u)$ at $0.2 < u < 0.8$ 
  depends almost entirely on  $\zeq$ and $n_s$, which
  are already well constrained by the CMB.  
  Therefore, we anticipate that 
   $W_b$ should depend mainly on $f_b$ and only weakly on $\zeq$. 

 To quantify this, we evaluated the ratio $W_b$ for our six models, 
  and the results are given in the
 last column of Table~\ref{tab:models}: the table shows 
  that $W_b$ is close to $0.61$ for all three $\neff = 3.04$ models,
  and close to $0.56$ for all three $\neff = 4.04$ models, 
  consistent with our expectations above.  
  The dependence of $W_b$ 
  on $\zeq$ is below 1\% and nearly negligible, while adding a 
 fourth neutrino species or equivalent 
  reduces $W_b$ by a factor close to $0.915$ 
 (i.e. 8.5\% suppression) in each case. 
  This reduction is slightly less than we would
  expect from linear scaling $W_b \propto f_b$, since our $\neff = 4.04$ 
 models have $f_b$ reduced by a factor $1.134^{-1} = 0.882$ relative to the
  corresponding $\neff = 3.04$ model.  
 The probable explanation is that baryons,
   in addition to causing oscillations, also affect the broad-band
 shape of the power spectrum \citep{eh98}  
 by suppressing power on all scales smaller than the sound horizon.
 Therefore, reducing the baryon fraction slightly increases power
  on intermediate scales, and changes the broad-band shape of $\xi(u)$, which 
   slightly counteracts the reduction in the bump area.

 The conclusion is that, {\em if} galaxy bias is scale-independent 
 on large scales and the area of the BAO peak is conserved under non-linear
 evolution (or can be recovered by reconstruction methods),  
 then measurements of $W_b$ can offer a potential new probe of $\neff$. 
 Estimates from large numerical simulations could be used to test these 
  assumptions, and possibly attempt to correct for
  any resulting biases.  

 {\newtwo The largest current redshift surveys  provide  
  a $\sim 6\sigma$ detection of the BAO peak \citep{anderson12}, 
  which would translate
 to approximately 16 percent uncertainty in $W_b$; this is twice
  as large as the 8.5 percent shift predicted above for $\neff \sim 4$, 
 so at present the precision on $\neff$ looks 
  uncompetitive with other methods. 
  However, future
  next-generation large redshift surveys can potentially offer
 a large improvement, and thus an interesting test of $\neff$
 which is complementary to the better-known methods from the CMB
  and nucleosynthesis. }

\subsection{Effect of Planck data} 
 \label{sec:planck} 

 Most of this paper studies models with parameter choices based on the
  WMAP-9 cosmological parameter results \citep{wmap9}; 
 the C3 model is near the
 best-fit, and L and H models have $\zeq$ shifted by 
  $\pm 1.5\sigma$ in WMAP units.   
 After this paper was nearly completed, the first {\em Planck} cosmology
 data release occurred in 2013 March.  While there
 are many interesting consequences for inflation and the early universe, 
  for the present purposes, two results are most notable: 
 firstly concerning $\neff$,  
 the evidence for $\neff > 3.04$ has generally 
 weakened \citep{planck-xvi}, 
 but the strength of this conclusion is somewhat dependent on 
 the choice of additional data sets. 
 
 The fit $\Lambda$CDM + varying $\neff$ to the dataset 
 ``{\em Planck} + WMAP polarisation + high-L + BAO'' 
 (the right column of Table~10 of \citealt{planck-xvi}) 
 gives $\neff = 3.30 \pm 0.26$, which 
 is $1\sigma$ above the standard value and 
  excludes $\neff = 4.04$ at the $2.8\sigma$ level. 
 However, there remains the well-publicised tension that {\em Planck}
  with vanilla 
  $\Lambda$CDM (and $\neff = 3.04$) prefers
  a value of $H_0 \approx 67.8 \pm 0.8 \hunit$, 
  which is below the $2\sigma$ range given by recent 
   local measurements (\citealt{riess11}; \citealt{freedman12}). 
  There are many possible explanations, but
  this $H_0$ tension can be 
  ameliorated by increasing $\neff$: e.g. fitting {\em Planck} 
  + $H_0$ data {\newtwo allowing variable $\neff$} 
  gives $H_0 = 72.1 \pm 1.9 \hunit$ and 
   $\neff = 3.62 \pm 0.25$, i.e.  
  $2.2\sigma$ above the standard $\neff$.  
  In summary, $\neff \sim 4.0$ is somewhat disfavoured by {\em Planck},
  but a value of $\neff \sim 3.5$ is completely 
  allowed or perhaps even preferred by combining all current data. 
  There are interesting possible models with extra relativistic species other
  than neutrinos  leading to 
   $\neff \sim 3.5$ (e.g. \citealt{wein13}). 

 Secondly, concerning $\zeq$ and $\Omega_{cb}$, 
  the {\em Planck} \, data imply values somewhat higher
  than WMAP; for the vanilla $\Lambda$CDM model, 
  fits to {\em Planck}+BAO data give $\zeq = 3366 \pm 39$ and 
  $\Omega_{cb} = 0.307 \pm 0.01$ (and $h = 0.678 \pm 0.008$
   for standard $\neff$).   
  The {\em Planck} constraints on $\zeq$ 
  are especially robust: in the many 
  extensions of $\Lambda$CDM considered by the {\em Planck} team, the
   bounds $3150 \le \zeq \le 3500$ are generic, i.e. values
   outside this range are excluded at $> 2\sigma$  
  for all of the added-parameter models and data combinations. 
 (Clearly, still more complicated models with even more non-vanilla
  parameters might widen this range; but there appears little motivation
  at present for adding two or more new parameters 
  beyond the basic six). 

 Comparing to our models above, the {\em Planck} central 
  value $\zeq \simeq 3366$ 
  is near the mid-point between our model pairs
  C and H above, but slightly closer to H.  Our two L models 
  ($\zeq = 3101$) are now firmly excluded by {\em Planck}, at 
  around the $5\sigma$ level for the base model or $3 \sigma$ for extended
   models.  
 Also, {\em Planck} prefers $\omega_b \simeq 0.0221$ which is just 2 percent
  below our default; and $n_s \simeq 0.961$ which is nearly identical.   
 Thus, while {\em Planck} has narrowed the allowed range of $\zeq$ and
  $\neff$, our models C3/C4/H3/H4 approximately bracket
   the range of $\zeq$ and $\neff$ allowed by {\em Planck};  
  and the main conclusions of this paper regarding the BAO amplitude
   are essentially unaffected.

\section{ Conclusions } 
\label{sec:conc} 

We have shown that a measurement of the BAO peak amplitude
 via the observable $W_b$ in Eq.~\ref{eq:wb} 
 may provide an interesting measurement of the cosmic baryon fraction; 
 this observable has been constructed so as to cancel galaxy bias, 
  non-linearity and dark energy effects to leading order, thus
 being sensitive mostly to $f_b$.  
 
 Comparing this BAO-based measurement to the measurement of 
  (approximately) $f_b \xrad$ from
 the CMB then gives an interesting probe of
  $\neff$;  this is largely {\newtwo complementary to} 
  the better-known method based on fitting the CMB damping tail.  
 Here, the key inputs required from the CMB are 
  constraints on $\zeq$ and $\omega_b$. 
 Assuming standard gravity and standard recombination, 
 these {\newtwo two parameters}   
 are very robust against extra-parameter extensions to vanilla $\Lambda$CDM. 

 There are two main assumptions used here:  
 firstly  that galaxy bias is nearly 
  scale-independent on the large scales between $30 < r < 120 \hmpc$, 
 and secondly that the area (not height) of the BAO bump is conserved 
 during the non-linear evolution of structure. 
 Both of these assumptions are reasonably well-motivated, but 
  much more detailed numerical simulations would be needed to see how
 well these approximations are expected to hold in practice. 

 A measurement of $W_b$ to useful precision 
  will require a substantial advance on 
  current data: the current precision on the BAO bump area is around
 16\%, while we would need to reach around $3\%$ to get a useful
 distinction between $\neff = 3$ or 4; this 
  appears a challenging proposition.  However, given 
 that the CMB temperature measurements are now approaching the 
  limits set by cosmic variance and foregrounds, 
  other independent probes of $\neff$ are
 highly desirable, and the test here should become feasible 
 at no extra cost from planned next-generation BAO redshift surveys.

\section*{Acknowledgements}

We thank the anonymous referee for helpful comments 
which significantly clarified this paper. 

We acknowledge the use of WMAP data from 
 the Legacy Archive for Microwave Background
 Data Analysis (LAMBDA) at GSFC (lambda.gsfc.nasa.gov),
  supported by the NASA Office of Space Science. 

 

\onecolumn 
\appendix
\section{Parameter dependence of CMB peaks and BAOs} 

 In this section we give some accurate approximations for  
  the dependence of CMB acoustic scale and
  BAO distance ratios on cosmological parameters, especially 
  $\zeq$ and $\Omega_{cb}$ used as basic parameters above.  
  This helps to understand the parameter choices 
  in Table~\ref{tab:models},  
 and the resulting shifts in BAO bump position observed in 
 \S~\ref{sec:camb}.   

 Firstly, we find that a very good approximation to the CMB acoustic
  wavenumber $\ell_*$ for models fairly close to standard $\Lambda$CDM is 
 \begin{equation}
 \label{eq:lstar} 
 \ell_* \simeq 301.9 \, \left( { 1 + \zeq \over 3201 } \right)^{-0.25}  
  \, \left( {\Omega_{cb} \over 0.270} \right)^{0.1} \, 
  \left( { 1 - f_\nu \over 0.995 } \right)^{0.4} 
  \, (1 + 1.6 \, \Omega_k) \, \left[1 - 0.11(1+w)\right]
\end{equation} 
 where $f_\nu \equiv \Omega_\nu/ (\Omega_{cb} + \Omega_\nu)$, and this 
  allows for small neutrino mass, weak curvature and 
   constant $w \ne -1$.  (This is for $\omega_b = 0.0226$; however,
  changing to the Planck value $\omega_b = 0.02215$ gives only around 
   0.1 percent reduction in $\ell_* $). 
 This has almost negligible dependence on $\neff$, since varying $\neff$
  (at fixed $\zeq , \Omega_{cb}$ as above) 
   results in both $r_s(z_*)$ and $D_A(z_*)$ 
  shrinking by a factor very close to $\xrad^{-0.5}$, but these cancel
  almost exactly in $\ell_*$. 
 
 Since $\ell_*$ is measured to  
  high precision $\approx 0.2$ percent 
  by WMAP+ACT+SPT,  if we vary $\zeq$ (as in the L/H models in 
  Table~\ref{tab:models} above), 
  then to remain consistent with CMB data we must adjust other parameter(s) 
  to preserve $\ell_*$.  
 Given our assumption
 of flat $\Lambda$ models and minimal neutrino mass in \S~\ref{sec:measfb}, 
 the only available parameter above is $\Omega_{cb}$. 
  Forcing a 5\% reduction in $\zeq$ (as chosen 
  for models L3/L4)  requires a 12\% reduction
  in $\Omega_{cb}$ to keep $\ell_*$ the same as the baseline model C3; 
 this corresponds to an increase in $h$ by 3.6\% (at fixed $\neff$). 
 Shifts from C to H models are basically the opposite of this. 
 {\newtwo We note one counter-intuitive feature: when varying parameters
  to conserve $\ell_*$,  
  it turns out that $h$ changes in the {\em opposite} sense to $\zeq$;  
  this is distinct from the common case of fixing $\Omega_{cb}$ and varying 
  $h$, when $1+\zeq$ varies $\propto h^2$. } 

 We can also understand the resulting shifts in the BAO bump location 
  as follows: 
 if we copy approximation (12) from \cite{suth12} for low-redshift
 BAO ratios, which is 
 \begin{equation} 
\label{eq:omez} 
  {z \, r_s \over D_V(z)} \simeq 0.01868 \, (1+\epsilon_V(z)) 
 { E(\frac{2}{3}z) \over \sqrt{\Omega_{cb}} }  
  \left( { 1 + \zeq \over 3201 } \right)^{0.25} \ ; 
\end{equation}  
  here the LHS is a direct observable from a BAO survey at
 effective redshift $z$,  
  $D_V(z)$ is the usual BAO dilation length \citep{eis05}, 
  $E(z) \equiv H(z)/H_0$, 
  and $\epsilon_V$ is a small cosmology-dependent 
  correction term \citep{suth12}, which is typically 
  $\simlt 0.05 z^2$ and effectively 
   negligible at modest redshift $z \simlt 0.3$.  
   Approximation~\ref{eq:omez} is accurate to $\simlt 0.7\%$
    at $z \le 0.4$, comparable to the cosmic variance limit, 
  and again this is almost independent of $\neff$.  
 At low redshift, Eq.~\ref{eq:omez} is only weakly sensitive
 to additional non-vanilla parameters such as curvature and varying $w$
  via the $E(2z/3)$ term; 
 this explains why low-redshift BAO observations provide a very
 robust constraint on $\Omega_{cb}$.

 In the limit $z \rightarrow 0$,  the above simplifies to 
 \begin{equation} 
\label{eq:rsh0} 
  { r_s \, H_0 \over c } \simeq \, { 0.01868 \over \sqrt{\Omega_{cb}} } 
    \, \left({1 + \zeq \over 3201}\right)^{0.25}  \ . 
\end{equation}  
  The LHS is equivalent to a hypothetical 
   BAO measurement at $z = 0$; this is not strictly observable 
 since cosmic variance prevents us measuring the BAO feature at
  $z \simlt 0.1$; but it is a modest extrapolation from real low-$z$ 
 BAO surveys. The main point is 
   since a galaxy redshift survey of course measures redshifts
    not distances,   
   the apparent BAO bump ``length'' presented in 
   $\hmpc$ units, as in Figure~\ref{fig:xi}, 
  is really measuring the ``BAO velocity'' $H_0 \, r_s$ 
   in units of $100 \, {\rm km \, s^{-1}}$.  
  Although this quantity contains $h$,
   in the case of varying $\neff$ this gets cancelled:  
   if we vary $\neff$ while holding fixed $\zeq$ (as appropriate for 
  fitting CMB data), then 
  $h$ and $r_s$ both depend on the radiation density as 
   $\xrad^{1/2}$ and $\xrad^{-1/2}$ respectively;  so 
    their product is almost independent of $\xrad$ and 
   only depends on the dimensionless parameters $\zeq$ and $\Omega_{cb}$, 
   plus a very weak dependence on $\omega_b$ which is negligible 
  at the current level of accuracy.   
  Therefore, the observed velocity scale of the BAO feature at low
   redshift is primarily measuring $\Omega_{cb}$, not $h$, which 
 explains why the BAO feature does not shift between the 3 and 4 neutrino
  model pairs in \S~\ref{sec:camb}.   

  Since all of the approximations above are nearly independent of
  $\neff$, this was the rationale for choosing $\zeq$ and $\Omega_{cb}$
  as two of the basic parameters: 
  observations of CMB and BAOs give
  us direct constraints on $\zeq$ and $\Omega_{cb}$, nearly 
  independent of $\neff$.  These two directly give a constraint on  
   $h / \sqrt{\xrad}$ from Eq.~\ref{eq:hzeq},  but give 
  almost no ability to measure $h$, $\xrad$ separately; this
  explains why WMAP+BAO alone currently have very 
  weak leverage on $\neff$, unless further dimensionful 
  data such as $H_0$ or $t_0$ is added. 

 Finally, the fact that $\Omega_{cb}$ appears with a $-0.5$ power 
 in Eq.~\ref{eq:rsh0} explains why the BAO feature shifts to smaller
  (larger) velocity scale for the models H (L) above. 
 
\bsp

\label{lastpage} 

\end{document}